\documentclass[journal]{IEEEtran}
\usepackage{amsmath}
\usepackage{amsmath,amsfonts,amsthm,bm} 
\usepackage{amssymb}
\usepackage{longtable}

\usepackage[table,xcdraw]{xcolor}

\usepackage{float}
\usepackage[ruled]{algorithm2e}

\usepackage{multicol}

\usepackage{caption}
\usepackage{graphicx}
\usepackage[noend]{algpseudocode}
\usepackage{subcaption}
\usepackage[noadjust]{cite}
\usepackage{tabularx}

\usepackage{fmtcount} 
\usepackage{array,multirow}
\usepackage{array, boldline, rotating}
\newcolumntype{?}{!{\vrule width 1pt}}

\usepackage{verbatim}

\begin{document}
%
%
\title{A Comprehensive Analysis of Machine Learning Models for Algorithmic Trading of Bitcoin}

\author{Abdul~Jabbar and Syed~Qaisar~Jalil
        
\thanks{Dr. A. Jabbar, and Dr. S.Q. Jalil are with Neurog LLP. E-mails: abduljabbar@neurog.ai, syedqaisarjalil@neurog.ai}

}

\maketitle

\begin{abstract}
This study evaluates the performance of 41 machine learning models, including 21 classifiers and 20 regressors, in predicting Bitcoin prices for algorithmic trading. By examining these models under various market conditions, we highlight their accuracy, robustness, and adaptability to the volatile cryptocurrency market. Our comprehensive analysis reveals the strengths and limitations of each model, providing critical insights for developing effective trading strategies. We employ both machine learning metrics (e.g., Mean Absolute Error, Root Mean Squared Error) and trading metrics (e.g., Profit and Loss percentage, Sharpe Ratio) to assess model performance. Our evaluation includes backtesting on historical data, forward testing on recent unseen data, and real-world trading scenarios, ensuring the robustness and practical applicability of our models. Key findings demonstrate that certain models, such as Random Forest and Stochastic Gradient Descent, outperform others in terms of profit and risk management. These insights offer valuable guidance for traders and researchers aiming to leverage machine learning for cryptocurrency trading.

\end{abstract}
\begin{IEEEkeywords}
Bitcoin, Machine Learning, Trading Strategies
\end{IEEEkeywords}

\IEEEpeerreviewmaketitle

\section{Introduction}
The advent of Bitcoin and the subsequent proliferation of cryptocurrencies have not only disrupted traditional financial systems but also introduced novel paradigms in asset trading. Cryptocurrencies, led by Bitcoin, have carved a niche in financial markets, attracting attention from both retail and institutional investors. The allure of high returns, coupled with the inherent volatility of these digital assets, has spurred the development of sophisticated trading strategies. Among these, algorithmic trading, leveraging the prowess of machine learning models, has emerged as a key player in navigating the cryptocurrency market landscape \cite{fang2022cryptocurrency}.

Bitcoin, the forerunner in this domain, presents a unique blend of challenges and opportunities for traders. Its decentralized nature, coupled with the absence of regulatory oversight, results in significant price fluctuations. This volatility, while posing risks, also creates opportunities for substantial gains, making Bitcoin an attractive asset for algorithmic trading strategies. These strategies, which were once the domain of sophisticated institutional traders, are now increasingly accessible to a wider audience, thanks to advancements in computational power and machine learning techniques.

The integration of machine learning in trading strategies for Bitcoin and other cryptocurrencies represents a significant shift from traditional trading approaches. Machine learning models offer the capability to process and learn from vast datasets, including historical price movements, trading volumes, and market sentiments. This ability to extract meaningful patterns and insights from complex and often noisy data is crucial in predicting future market behavior and making informed trading decisions \cite{dakalbab2024artificial}.

Our research delves deep into the realm of algorithmic trading for Bitcoin, employing a range of machine learning models. The primary aim is to critically analyze the performance of these models in the context of Bitcoin trading. We explore various aspects of these models, including their predictive accuracy, response to market volatility, and the effectiveness of different feature sets. In doing so, this study sheds light on the nuances of algorithmic trading in the cryptocurrency market and provides a roadmap for traders and investors in navigating this volatile yet potentially lucrative domain.

A key motivation behind this study is the growing interest in cryptocurrency trading and the need for robust trading strategies that can adapt to the dynamic nature of these markets. The extreme volatility of cryptocurrencies, while a deterrent for some, presents a fertile ground for algorithmic trading strategies. Machine learning models, with their adaptability and learning capabilities, are well-suited to capture the intricacies of these markets. However, it is imperative to critically assess the performance of different machine learning models, each with its unique strengths and limitations, to identify the most effective strategies for cryptocurrency trading.

Our research contributes to the existing literature by providing a comprehensive analysis of various machine learning models in the context of Bitcoin trading. We not only assess the performance of these models but also explore the implications of their results in practical trading scenarios. This includes considerations of market volatility, transaction costs, and other factors that impact on the trading outcomes.

The structure of the paper is as follows: In Section 2, we present a detailed literature review, examining previous work in this field and identifying gaps that our study aims to fill. Section 3 describes the methodology, including the data sources, machine learning models employed, and the evaluation criteria used. Section 4 discusses the results, providing an in-depth analysis of the performance of each model. Section 5 offers a discussion on the implications of our findings, both for traders and the broader field of financial machine learning. Finally, Section 6 concludes the paper, summarizing our key findings and suggesting avenues for future research.

Through this comprehensive exploration, our study aims to not only advance the understanding of algorithmic trading in the cryptocurrency sphere but also to provide practical insights that can be leveraged by traders and investors.

\section{Literature Review} 
The application of machine learning (ML) techniques to predict cryptocurrency prices has garnered significant attention due to the volatile nature of these markets and the potential for substantial financial returns. This section reviews key studies in this domain, highlighting methodologies, findings, and how our study advances the current state of knowledge.

Machine learning algorithms have been extensively employed to predict Bitcoin prices, leveraging their ability to handle large datasets and capture complex patterns. Several studies have explored various ML models and techniques to enhance prediction accuracy. For example, \cite{nagamani2023bitcoin} applied Support Vector Machine (SVM) and K-Nearest Neighbor (KNN) algorithms to forecast Bitcoin prices, demonstrating that SVM outperforms KNN in terms of accuracy. This study emphasizes the importance of machine learning in producing more accurate results compared to traditional techniques. Similarly, \cite{maleki2020bitcoin} investigated the prediction of Bitcoin prices using the prices of other cryptocurrencies, such as Ethereum, Zcash, and Litecoin. They employed cointegration analysis, regression models, and ARIMA models to analyze price trends and found that Zcash performed best in forecasting Bitcoin prices without direct Bitcoin price information.

Highlighting the superiority of machine learning over traditional methods, \cite{akyildirim2021forecasting} evaluated the forecasting performance of various ML algorithms using high-frequency intraday data. They found that SVM achieved the highest accuracy, outperforming traditional models like ARIMA, especially during market turmoil such as the COVID-19 pandemic. In a different approach, \cite{ziweritin2023height} combined a high-end multi-layer perceptron (MLP) with various machine learning techniques to predict Bitcoin prices. This study achieved high prediction accuracies using optimization techniques and classifiers like KNN and SVM.

Several studies have conducted comparative analyses of different ML models to identify the most effective techniques for cryptocurrency price prediction. \cite{siva2024analyzing} analyzed various machine learning methods for predicting Bitcoin prices, highlighting the superior prediction accuracy of Artificial Neural Networks (ANN) and SVMs compared to traditional parametric regression approaches. Additionally, \cite{al2023empirical} evaluated SVM, KNN, and Light Gradient Boosted Machine (LGBM) in predicting price movements of Bitcoin, Ethereum, and Litecoin. They found that KNN outperformed other models in the overall dataset, while SVM and LGBM were better for specific cryptocurrencies. Supporting these findings, \cite{yudono2022bitcoin} compared the effectiveness of Simple Moving Average (SMA) and Radial Basis Function Neural Network (RBFNN) methods. The study demonstrated that RBFNN significantly outperforms SMA, providing a more accurate tool for forecasting Bitcoin prices.

Advanced machine learning techniques, including ensemble methods, have shown promising results in predicting cryptocurrency prices. \cite{sebastiao2021forecasting} explored the predictability of major cryptocurrencies using linear models, random forests, and SVMs. The study found that ensemble approaches achieve significant profitability, particularly during bear market periods. Additionally, \cite{gyamerah2019bitcoins} investigated the predictability of Bitcoin prices using a stacking ensemble model, integrating Random Forest and Generalized Linear Model with Support Vector Regression (SVR) as a meta-learner. The study achieved high predictive accuracy, suggesting the effectiveness of ensemble methods.

Studies have also focused on practical applications and real-world testing of ML models to validate their performance and applicability in actual trading scenarios. \cite{yaoprediction} applied various ML techniques, including Logistic Regression, SVM, Random Forest, XGBoost, and LightGBM, to predict Bitcoin price movements. The study highlighted the potential of ensemble models in enhancing prediction accuracy and constructing effective trading strategies. Similarly, \cite{kumar2021comparative} conducted a comparative analysis of ARIMA, Facebook Prophet, and XGBoost to predict the monthly Bitcoin price rate. The results indicated that Facebook Prophet outperformed the other models, demonstrating high accuracy and reliability. Furthermore, \cite{falcon2021daily} performed a comparative analysis of machine learning models for forecasting next-day cryptocurrency returns. They found that SVMs provided the highest classification accuracy and developed a probability-based trading strategy that significantly outperformed standalone investments.

Some studies have integrated sentiment analysis and technical indicators to improve the accuracy of cryptocurrency price predictions. For instance, \cite{valencia2019price} applied machine learning and sentiment analysis techniques to predict price movements of major cryptocurrencies. The study leveraged data from Twitter and market data, finding that neural networks outperformed other models. Additionally, \cite{parra2023your} investigated the application of ML algorithms to forecast Bitcoin price movements. The study found that Random Forest achieved the highest forecasting performance on continuous datasets, while ANN performed best on discrete datasets.

Various performance metrics have been used to evaluate the effectiveness of ML models in predicting cryptocurrency prices. \cite{iqbal2021time} compared ARIMA, Facebook Prophet, and XGBoost using metrics such as RMSE, MAE, and R-squared. The study demonstrated that ARIMA outperformed the other models, highlighting the importance of preprocessing and feature selection. Similarly, \cite{septiarini2020comparative} investigated the efficacy of ML algorithms in predicting Bitcoin prices. The study found that RF exhibited the highest forecasting accuracy on continuous datasets, while ANN performed best on discrete datasets.

While existing studies have significantly advanced the field of cryptocurrency price prediction, they often face challenges related to model robustness, overfitting, and the ability to adapt to rapidly changing market conditions. Our study addresses these challenges by integrating both machine learning and trading metrics (e.g., Mean Absolute Error, Root Mean Squared Error, Profit and Loss percentage, Sharpe Ratio) to comprehensively evaluate model performance. Furthermore, our evaluation process includes backtesting on historical data, forward testing on recent unseen data, and real-world testing to ensure robustness and practical applicability. This multi-faceted evaluation approach provides a more thorough assessment of model performance compared to previous studies.

Key findings from our study demonstrate that certain models, such as Random Forest and Stochastic Gradient Descent, outperform others in terms of profit and risk management. These insights offer valuable guidance for traders and researchers aiming to leverage machine learning for cryptocurrency trading, highlighting the practical benefits and improved accuracy of our approach. By incorporating economic indicators and considering practical trading constraints, our study contributes to the development of more efficient and reliable algorithmic trading strategies in the cryptocurrency domain. This comprehensive evaluation framework and the integration of diverse metrics set our study apart from previous research, offering a more robust and practical solution for Bitcoin price prediction and trading.

\section{Methodology}
\subsection{Data}
In machine learning, the quality and depth of data are critical, especially in complex fields like financial trading. For Bitcoin trading, the challenge is even more pronounced due to the market's relatively recent development and the lack of centralized, comprehensive historical data. To circumvent these challenges, our study leverages a detailed dataset of Bitcoin prices, publicly available since the inception of Bitcoin trading in 2013. This extensive dataset is invaluable for training models capable of recognizing and adapting to a wide spectrum of market conditions, which is essential for developing sophisticated algorithmic trading strategies.

The dataset for this research is meticulously divided into three segments: training, backtesting, and forward testing. The training dataset spans a decade, from January 2013 to January 2023, providing a rich historical context for the models to learn from. This lengthy period is crucial to encompass the diverse range of market behaviors and trends Bitcoin has experienced. The backtesting phase covers six months, from February to July 2023, and is instrumental in evaluating the models on unseen data, thus testing their ability to generalize beyond the training set. This is a crucial step in preventing overfitting. Finally, the forward testing phase, from August to October 2023, serves as a real-world application of the models, ensuring they are tested against new, unencountered data, thereby eliminating any survivorship bias.

\begin{figure*}[ht]
\centering
\includegraphics[width=0.8\textwidth]{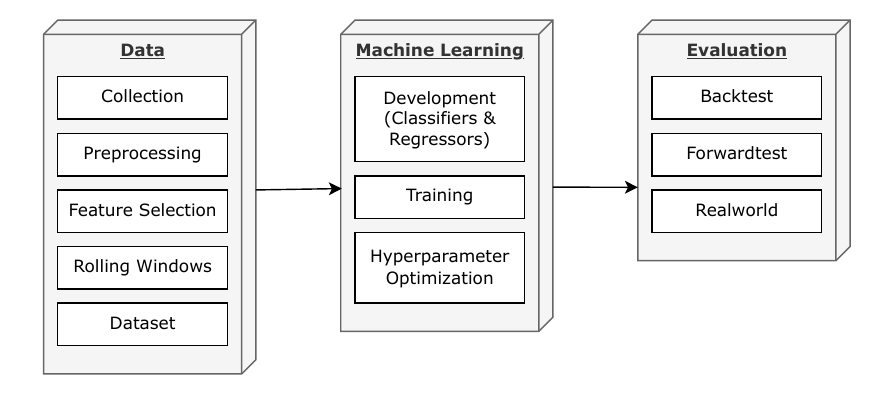}
\caption{Overview of the Methodology: This flowchart illustrates the comprehensive process used in our study, encompassing three main modules: data, machine learning, and evaluation. The data module includes all the steps from data collection to dataset creation, preparing the data for use by the machine learning module. The machine learning module covers model development and training, including hyperparameter optimization for both classifiers and regressors. The evaluation module involves rigorous backtesting on historical data, forward testing on recent unseen data, and real-world testing to validate model performance and ensure practical applicability.}
\label{fig:methodology_flowchart}
\end{figure*}

To enhance the models' input features, the study incorporates a range of technical indicators alongside the raw pricing data. These indicators include:
\begin{itemize}
    \item \textbf{Accumulation/Distribution Index}: A volume-based indicator designed to reflect cumulative inflows and outflows of money, providing insights into the strength of a trend based on volume movements.
    \item \textbf{Money Flow Index (MFI)}: This indicator combines price and volume to identify overbought or oversold conditions in an asset, offering a perspective on the intensity of buying or selling pressure.
    \item \textbf{Bollinger Bands}: A statistical chart characterizing the prices and volatility of an asset over time, which includes a moving average and two standard deviation lines.
    \item \textbf{Keltner Channel Width}: This encompasses a volatility-based envelope set above and below an exponential moving average of the price, offering insights into potential trend breakouts or reversals.
    \item \textbf{Parabolic SAR (Stop and Reverse)}: This indicator is used to determine the direction of an asset's momentum and the point in time when this momentum has a higher-than-normal probability of switching directions.
\end{itemize}

Each of these indicators provides a unique lens through which to analyze market trends and movements, and their incorporation is expected to enrich the feature set available for our machine learning models.

Our methodology is further characterized by the use of rolling windows of various sizes: 1, 7, 14, 21, and 28 days. This approach ensures that our models have access to a dynamic, evolving view of market conditions, as each window encompasses the preceding $n$ intervals of data. Such a technique is crucial for models that need to understand and predict market trends over different time horizons. The models that show the highest performance, particularly in terms of profit and loss (PNL) percentage across these windows, are then selected for detailed examination in the subsequent sections of this study.

An important preprocessing step applied to our dataset is the log difference transformation. Mathematically, this can be expressed as:
\[ \Delta \log(P_t) = \log(P_t) - \log(P_{t-1}) \]
where \(P_t\) and \(P_{t-1}\) represent the price of Bitcoin at times \(t\) and \(t-1\), respectively. This transformation is effective in stabilizing variance, linearizing trends, and introducing stationarity to the dataset, crucial for analyzing financial time series where understanding growth rates and temporal changes is important.

Finally, the design of our dataset is intentionally made flexible to accommodate various time intervals. While the primary focus is on a 24-hour trading horizon, the structure is adaptable to different temporal scales. This flexibility showcases the broad applicability of our methodology, suitable for a range of trading frequencies and market conditions.

\subsection{Machine Learning Models}
In this research, a diverse array of machine learning classifiers and regressors has been employed to analyze and predict Bitcoin trading patterns. Each model has been meticulously selected for its unique attributes and potential effectiveness in capturing the complexities of the cryptocurrency market. Classifiers are tasked with determining the trading action—specifically, whether to buy (go long) or sell (go short). In contrast, regressors focus on predicting the magnitude of price changes over specified intervals. To distinguish between the two, we denote classifiers with a suffix 'C' and regressors with 'R'.

\subsubsection{Classifiers}
The following classifiers have been employed:

\begin{enumerate}

    \item \textbf{Ada Boost (ABC)}: This ensemble method combines multiple weak learners to form a stronger model, enhancing performance in varied market conditions.
    \item \textbf{Bagging (BGC)}: Uses bootstrap aggregating to improve stability and reduce overfitting, crucial in volatile market scenarios.
    \item \textbf{Bernoulli NB (BNBC)}: Suited for binary classification, it's effective in scenarios with binary/boolean feature sets.
    \item \textbf{Calibrated CV (CCVC)}: Improves probability estimation in classification, essential for better trade decision-making.
    \item \textbf{Decision Tree (DTC)}: Offers a transparent, tree-structured modeling approach, useful for clear interpretation of trading signals.
    \item \textbf{Extra Tree (ETC)}: A Random Forest variant that introduces more randomness in split decisions, aiming to reduce model overfitting.
    \item \textbf{Gaussian Process (GPC)}: Excellent for small datasets, captures complex patterns using kernel functions, suitable for nuanced market analysis.
    \item \textbf{K Neighbors (KNC)}: A non-parametric method that classifies based on the proximity to nearest neighbors, useful in identifying market trends.
    \item \textbf{Linear Discriminant Analysis (LDAC)}: Effective in finding linear combinations of features for class separation, suitable for linearly separable market data.
    \item \textbf{Linear SVC (LSVC)}: Applies Support Vector Classification in scenarios with linear separability, efficient for clear market trend data.
    \item \textbf{Logistic Regression (LRC)}: A fundamental model for binary classification, ideal for straightforward buy or sell decisions.
    \item \textbf{Logistic Regression CV (LRCVC)}: Integrates logistic regression with cross-validation, optimizing for the best model parameters.
    \item \textbf{MLP (MLPC)}: A neural network-based model, capable of capturing complex, non-linear relationships in market data.
    \item \textbf{Passive Aggressive (PAC)}: Suitable for large-scale learning, it updates models based on prediction errors, adapting swiftly to market changes.
    \item \textbf{Perceptron (PC)}: A simple, yet effective linear classifier for large datasets, efficient in handling vast market data.
    \item \textbf{Quadratic Discriminant Analysis (QDAC)}: Assumes Gaussian distribution for class separation, effective in markets exhibiting normal distribution patterns.
    \item \textbf{Random Forest (RFC)}: An ensemble of decision trees, known for high accuracy and robustness against overfitting in complex market environments.
    \item \textbf{Ridge (RC)}: A linear model with L2 regularization, adept at handling multicollinearity in financial datasets.
    \item \textbf{SGD (SGDC)}: Utilizes stochastic gradient descent for optimized computational efficiency, crucial in high-frequency trading scenarios.
    \item \textbf{SVC (SVC)}: Versatile in handling both non-linear and high-dimensional data, adaptable to various market conditions.
    \item \textbf{Radius Neighbors (RNC)}: Classifies based on a fixed radius, useful in spatial or locality-based market analyses.
\end{enumerate}

\subsubsection{Regressors}
The regressors used in this study are as follows:

\begin{enumerate}
    \item \textbf{Ada Boost (ABR)}: Applies an ensemble technique focusing on challenging data points, enhancing accuracy in regression tasks.
    \item \textbf{Bagging (BGR)}: Employs bootstrap sampling to create multiple models, reducing variance and improving predictions in regression.
    \item \textbf{Decision Tree (DTR)}: An interpretable model for regression, useful in capturing non-linear relationships in price movements.
    \item \textbf{Extra Tree (ETR)}: Improves on Random Forest by randomizing decision trees, enhancing regression performance in unpredictable markets.
    \item \textbf{Gaussian Process (GPR)}: Ideal for small datasets with complex patterns, offers probabilistic outputs beneficial for risk assessment.
    \item \textbf{K Neighbors (KNR)}: Predicts values based on the proximity of neighbors, effective in markets with spatial correlation.
    \item \textbf{Linear SVR (LSVR)}: Adapts Support Vector Regression for linear contexts, efficient in markets with linear price movements.
    \item \textbf{MLP (MLPR)}: A neural network approach for modeling complex regression patterns in financial markets.
    \item \textbf{Random Forest (RFR)}: Known for high accuracy in regression, leveraging an ensemble of decision trees to predict price changes.
    \item \textbf{Ridge (RR)}: Utilizes L2 regularization to mitigate overfitting in regression, essential for stable financial predictions.
    \item \textbf{SGD (SGDR)}: Implements stochastic gradient descent for efficient regression analysis in large datasets.
    \item \textbf{SVR (SVRR)}: A versatile kernel-based method, effective for both linear and non-linear regression tasks in trading.
    \item \textbf{ARD (ARDR)}: Uses Automatic Relevance Determination to adapt regression models to the inherent structure of the data.
    \item \textbf{Bayesian Ridge (BRR)}: Combines ridge regression with Bayesian inference, offering flexible modeling in uncertain market conditions.
    \item \textbf{Gradient Boosting (GBR)}: Constructs an additive model in a forward stage-wise fashion, useful in progressive market trend analysis.
    \item \textbf{Lars (LaR)}: Efficient in high-dimensional data regression, providing solutions along a regularization path.
    \item \textbf{Linear Regression (LiR)}: The foundational regression model, establishing linear relationships between market variables.
    \item \textbf{RANSAC (RanR)}: Fits models robustly to subsets of data, effectively dealing with outliers in financial datasets.
    \item \textbf{Theil Sen (TSR)}: A non-parametric approach resilient to outliers, suitable for complex multivariate regression in trading.
    \item \textbf{Radius Neighbors (RNR)}: Utilizes a fixed radius for neighborhood-based regression, applicable in spatially correlated market environments.
\end{enumerate}

The selection of these diverse models is based on their established effectiveness in predictive modeling, particularly in the financial markets where accuracy, adaptability, and robustness are of utmost importance. This wide range of models ensures a comprehensive analysis, allowing us to identify the most effective strategies for Bitcoin trading prediction.

\subsection{Rolling Windows and Training Process}
The concept of rolling windows is pivotal in time series analysis, especially in financial markets where data is sequential and market conditions are dynamic. A rolling window approach involves using a $window$ of a fixed size that moves through the dataset over time. For each position of the window, a subset of data is selected, which is then used for training the model. This technique is crucial in capturing the evolving nature of financial markets, as it allows models to learn from the most recent trends and patterns.

In the context of machine learning for Bitcoin trading, rolling windows are essential for several reasons. Firstly, they enable models to adapt to changing market conditions, which is crucial in a volatile market like Bitcoin. By training on the most recent data, the models stay updated with current market dynamics, enhancing their predictive accuracy. Secondly, rolling windows help in mitigating the risk of overfitting. Models trained on a specific period might perform well on that period but fail to generalize to new data. By continuously updating the training dataset, rolling windows ensure that models are not overly tuned to a specific historical period.

In this study, five different rolling window sizes were used: 1, 7, 14, 21, and 28 days. These sizes were chosen to capture various market dynamics, from short-term fluctuations to longer-term trends. Each window size provides a different perspective on the data, allowing models to learn patterns and trends over different time horizons. For instance, a 1-day window focuses on very short-term movements, while a 28-day window captures broader market trends.

Each machine learning model in our study was trained against each rolling window size. This process involved sequentially moving the window through the entire dataset, training the model on the data within the window at each step. For example, with a 7-day window, the model would be trained on data from days 1 to 7, then on data from days 2 to 8, and so on, until the end of the dataset. This approach ensures that each model is exposed to a wide range of market conditions, enhancing its ability to generalize and adapt.

The use of multiple window sizes allows us to analyze the performance of each model under different market conditions. It provides insights into which models are better at capturing short-term trends versus long-term trends. This is particularly important in Bitcoin trading, where market conditions can change rapidly. Models that perform well across multiple window sizes are likely to be more robust and versatile, making them more reliable for real-world trading applications.

After training, each model's performance was evaluated based on its predictive accuracy within each window. The model with the highest performance in terms of predictive accuracy and profitability (PNL) for each window size was then selected for further analysis. This approach allows us to identify the most effective models for Bitcoin trading, considering both short-term and long-term market behaviors.

\subsection{Hyperparameter Optimization}
Hyperparameters are the configurable settings used to tune the performance of machine learning models. Unlike model parameters, which are learned during training, hyperparameters are set prior to the training process and can have a significant impact on the effectiveness of the models. Proper hyperparameter optimization is critical in machine learning, particularly in financial applications like Bitcoin trading, where the optimal model configuration can substantially influence predictive accuracy and profitability.

For the purpose of hyperparameter optimization in this study, we employed Optuna \cite{optuna_2019}, an open-source hyperparameter optimization framework. Optuna is designed for automating the process of finding the best hyperparameters, making it an ideal tool for our complex machine learning tasks. It uses a Bayesian optimization technique to search the hyperparameter space efficiently, focusing on combinations that are more likely to yield better model performance. This approach is especially beneficial given the large number of models and the extensive range of hyperparameters involved in our study.

In our implementation with Optuna, each model underwent 100 trials of hyperparameter tuning. In each trial, Optuna varied the hyperparameters within predefined ranges, searching for the combination that maximized the model's performance. The hyperparameters varied included learning rates, regularization strengths, the number of layers and neurons in neural network models, and other model-specific parameters. The variation in these hyperparameters was guided by Optuna's optimization algorithm, which adapted its search strategy based on the results of previous trials, thereby progressively honing in on the most promising hyperparameter values.

The primary metric for evaluating the performance of the models during the hyperparameter optimization process was the Profit and Loss (PNL) percentage. PNL was chosen as it directly reflects the financial efficacy of the models in trading scenarios. For each model, the hyperparameter combination that yielded the highest PNL percentage during the backtesting phase was identified as the optimal set. This approach ensured that the selected hyperparameters were not only statistically effective but also financially practical in terms of trading performance.

The optimization of hyperparameters is particularly important in the volatile and unpredictable domain of Bitcoin trading. Bitcoin markets exhibit unique characteristics and can behave differently from traditional financial markets. Therefore, fine-tuning the models to adapt to these idiosyncrasies through hyperparameter optimization is essential to achieve the best possible predictive performance.

\subsection{Backtest and Forward Test Procedures}
In financial machine learning applications, backtesting and forward testing are crucial steps for evaluating the effectiveness and robustness of models. Backtesting involves testing the models against historical data to assess their performance, while forward testing (also known as paper trading) tests the models on more recent, unseen data to evaluate how well they might perform in real-world trading scenarios.

For the purpose of this research, the dataset was divided into three distinct segments: training, backtesting, and forward testing. The training set, spanning from January 2013 to January 2023, was used to train the models. The backtesting phase covered data from February to July 2023, providing a recent historical dataset to evaluate the trained models. The forward testing phase, encompassing data from August to October 2023, served as a real-world test bed to assess the models' performance on new, unseen data.

\renewcommand{\arraystretch}{1.5}

\begin{table*}[]
\centering
\resizebox{\textwidth}{!}{%
\begin{tabular}{cccccccccccccccccc}
\hline
\multicolumn{1}{|c|}{} & \multicolumn{1}{c|}{} & \multicolumn{8}{c|}{\textbf{Backtest}} & \multicolumn{8}{c|}{\textbf{Forwardtest}} \\ \cline{3-18} 
\multicolumn{1}{|c|}{\multirow{-2}{*}{\textbf{Classifier}}} & \multicolumn{1}{c|}{\multirow{-2}{*}{\textbf{\begin{tabular}[c]{@{}c@{}}Rolling \\ window\end{tabular}}}} & \multicolumn{1}{c|}{\textbf{\begin{tabular}[c]{@{}c@{}}PNL \\ (\%)\end{tabular}}} & \multicolumn{1}{c|}{\textbf{Sharpe}} & \multicolumn{1}{c|}{\textbf{R2}} & \multicolumn{1}{c|}{\textbf{Accuracy}} & \multicolumn{1}{c|}{\textbf{\begin{tabular}[c]{@{}c@{}}F1 \\ score\end{tabular}}} & \multicolumn{1}{c|}{\textbf{Precision}} & \multicolumn{1}{c|}{\textbf{Recall}} & \multicolumn{1}{c|}{\textbf{\begin{tabular}[c]{@{}c@{}}No. of \\ Trades\end{tabular}}} & \multicolumn{1}{c|}{\textbf{\begin{tabular}[c]{@{}c@{}}PNL \\ (\%)\end{tabular}}} & \multicolumn{1}{c|}{\textbf{Sharpe}} & \multicolumn{1}{c|}{\textbf{R2}} & \multicolumn{1}{c|}{\textbf{Accuracy}} & \multicolumn{1}{c|}{\textbf{\begin{tabular}[c]{@{}c@{}}F1 \\ score\end{tabular}}} & \multicolumn{1}{c|}{\textbf{Precision}} & \multicolumn{1}{c|}{\textbf{Recall}} & \multicolumn{1}{c|}{\textbf{\begin{tabular}[c]{@{}c@{}}No. of \\ Trades\end{tabular}}} \\ \hline
\multicolumn{1}{|c|}{AdaBoostClassifier} & \multicolumn{1}{c|}{21} & \multicolumn{1}{c|}{89.26} & \multicolumn{1}{c|}{6.47} & \multicolumn{1}{c|}{0.92} & \multicolumn{1}{c|}{0.53} & \multicolumn{1}{c|}{0.64} & \multicolumn{1}{c|}{0.51} & \multicolumn{1}{c|}{0.88} & \multicolumn{1}{c|}{55} & \multicolumn{1}{c|}{-9.24} & \multicolumn{1}{c|}{-0.81} & \multicolumn{1}{c|}{0} & \multicolumn{1}{c|}{0.49} & \multicolumn{1}{c|}{0.59} & \multicolumn{1}{c|}{0.48} & \multicolumn{1}{c|}{0.77} & \multicolumn{1}{c|}{40} \\ \hline
\multicolumn{1}{|c|}{BaggingClassifier} & \multicolumn{1}{c|}{28} & \multicolumn{1}{c|}{\cellcolor[HTML]{FFCCC9}\textbf{121.73}} & \multicolumn{1}{c|}{\cellcolor[HTML]{FFCCC9}\textbf{7.17}} & \multicolumn{1}{c|}{\cellcolor[HTML]{FFCCC9}\textbf{0.89}} & \multicolumn{1}{c|}{\cellcolor[HTML]{FFCCC9}\textbf{0.6}} & \multicolumn{1}{c|}{\cellcolor[HTML]{FFCCC9}\textbf{0.65}} & \multicolumn{1}{c|}{\cellcolor[HTML]{FFCCC9}\textbf{0.57}} & \multicolumn{1}{c|}{\cellcolor[HTML]{FFCCC9}\textbf{0.74}} & \multicolumn{1}{c|}{\cellcolor[HTML]{FFCCC9}\textbf{74}} & \multicolumn{1}{c|}{-21.67} & \multicolumn{1}{c|}{-2.78} & \multicolumn{1}{c|}{0.5} & \multicolumn{1}{c|}{0.53} & \multicolumn{1}{c|}{0.59} & \multicolumn{1}{c|}{0.51} & \multicolumn{1}{c|}{0.7} & \multicolumn{1}{c|}{40} \\ \hline
\multicolumn{1}{|c|}{BernoulliNB} & \multicolumn{1}{c|}{21} & \multicolumn{1}{c|}{\cellcolor[HTML]{FFCCC9}\textbf{113.31}} & \multicolumn{1}{c|}{\cellcolor[HTML]{FFCCC9}\textbf{6.14}} & \multicolumn{1}{c|}{\cellcolor[HTML]{FFCCC9}\textbf{0.89}} & \multicolumn{1}{c|}{\cellcolor[HTML]{FFCCC9}\textbf{0.58}} & \multicolumn{1}{c|}{\cellcolor[HTML]{FFCCC9}\textbf{0.59}} & \multicolumn{1}{c|}{\cellcolor[HTML]{FFCCC9}\textbf{0.56}} & \multicolumn{1}{c|}{\cellcolor[HTML]{FFCCC9}\textbf{0.63}} & \multicolumn{1}{c|}{\cellcolor[HTML]{FFCCC9}\textbf{106}} & \multicolumn{1}{c|}{\cellcolor[HTML]{9AFF99}\textbf{29.78}} & \multicolumn{1}{c|}{\cellcolor[HTML]{9AFF99}\textbf{5.17}} & \multicolumn{1}{c|}{\cellcolor[HTML]{9AFF99}\textbf{0.84}} & \multicolumn{1}{c|}{\cellcolor[HTML]{9AFF99}\textbf{0.52}} & \multicolumn{1}{c|}{\cellcolor[HTML]{9AFF99}\textbf{0.53}} & \multicolumn{1}{c|}{\cellcolor[HTML]{9AFF99}\textbf{0.5}} & \multicolumn{1}{c|}{\cellcolor[HTML]{9AFF99}\textbf{0.56}} & \multicolumn{1}{c|}{\cellcolor[HTML]{9AFF99}\textbf{38}} \\ \hline
\multicolumn{1}{|c|}{CalibratedClassifierCV} & \multicolumn{1}{c|}{28} & \multicolumn{1}{c|}{92} & \multicolumn{1}{c|}{7.59} & \multicolumn{1}{c|}{0.86} & \multicolumn{1}{c|}{0.52} & \multicolumn{1}{c|}{0.66} & \multicolumn{1}{c|}{0.51} & \multicolumn{1}{c|}{0.94} & \multicolumn{1}{c|}{24} & \multicolumn{1}{c|}{-2.78} & \multicolumn{1}{c|}{0.05} & \multicolumn{1}{c|}{0.11} & \multicolumn{1}{c|}{0.42} & \multicolumn{1}{c|}{0.54} & \multicolumn{1}{c|}{0.44} & \multicolumn{1}{c|}{0.72} & \multicolumn{1}{c|}{30} \\ \hline
\multicolumn{1}{|c|}{DecisionTreeClassifier} & \multicolumn{1}{c|}{28} & \multicolumn{1}{c|}{62.22} & \multicolumn{1}{c|}{3.81} & \multicolumn{1}{c|}{0.62} & \multicolumn{1}{c|}{0.5} & \multicolumn{1}{c|}{0.63} & \multicolumn{1}{c|}{0.5} & \multicolumn{1}{c|}{0.88} & \multicolumn{1}{c|}{41} & \multicolumn{1}{c|}{-8.17} & \multicolumn{1}{c|}{-1.12} & \multicolumn{1}{c|}{0.28} & \multicolumn{1}{c|}{0.44} & \multicolumn{1}{c|}{0.57} & \multicolumn{1}{c|}{0.45} & \multicolumn{1}{c|}{0.77} & \multicolumn{1}{c|}{28} \\ \hline
\multicolumn{1}{|c|}{ExtraTreeClassifier} & \multicolumn{1}{c|}{28} & \multicolumn{1}{c|}{103.73} & \multicolumn{1}{c|}{6.14} & \multicolumn{1}{c|}{0.9} & \multicolumn{1}{c|}{0.57} & \multicolumn{1}{c|}{0.65} & \multicolumn{1}{c|}{0.54} & \multicolumn{1}{c|}{0.82} & \multicolumn{1}{c|}{49} & \multicolumn{1}{c|}{12.16} & \multicolumn{1}{c|}{2.46} & \multicolumn{1}{c|}{0.22} & \multicolumn{1}{c|}{0.49} & \multicolumn{1}{c|}{0.58} & \multicolumn{1}{c|}{0.48} & \multicolumn{1}{c|}{0.74} & \multicolumn{1}{c|}{31} \\ \hline
\multicolumn{1}{|c|}{GaussianProcessClassifier} & \multicolumn{1}{c|}{21} & \multicolumn{1}{c|}{47.36} & \multicolumn{1}{c|}{3.42} & \multicolumn{1}{c|}{0.57} & \multicolumn{1}{c|}{0.52} & \multicolumn{1}{c|}{0.58} & \multicolumn{1}{c|}{0.5} & \multicolumn{1}{c|}{0.69} & \multicolumn{1}{c|}{90} & \multicolumn{1}{c|}{-24.72} & \multicolumn{1}{c|}{-3.37} & \multicolumn{1}{c|}{0.52} & \multicolumn{1}{c|}{0.43} & \multicolumn{1}{c|}{0.46} & \multicolumn{1}{c|}{0.42} & \multicolumn{1}{c|}{0.51} & \multicolumn{1}{c|}{40} \\ \hline
\multicolumn{1}{|c|}{KNeighborsClassifier} & \multicolumn{1}{c|}{28} & \multicolumn{1}{c|}{103.84} & \multicolumn{1}{c|}{5.44} & \multicolumn{1}{c|}{0.96} & \multicolumn{1}{c|}{0.56} & \multicolumn{1}{c|}{0.57} & \multicolumn{1}{c|}{0.55} & \multicolumn{1}{c|}{0.61} & \multicolumn{1}{c|}{95} & \multicolumn{1}{c|}{-5.14} & \multicolumn{1}{c|}{0} & \multicolumn{1}{c|}{0.57} & \multicolumn{1}{c|}{0.49} & \multicolumn{1}{c|}{0.47} & \multicolumn{1}{c|}{0.47} & \multicolumn{1}{c|}{0.47} & \multicolumn{1}{c|}{51} \\ \hline
\multicolumn{1}{|c|}{LinearDiscriminantAnalysis} & \multicolumn{1}{c|}{28} & \multicolumn{1}{c|}{88.7} & \multicolumn{1}{c|}{4.65} & \multicolumn{1}{c|}{0.85} & \multicolumn{1}{c|}{0.51} & \multicolumn{1}{c|}{0.58} & \multicolumn{1}{c|}{0.5} & \multicolumn{1}{c|}{0.67} & \multicolumn{1}{c|}{82} & \multicolumn{1}{c|}{-6.84} & \multicolumn{1}{c|}{-0.23} & \multicolumn{1}{c|}{0.24} & \multicolumn{1}{c|}{0.5} & \multicolumn{1}{c|}{0.54} & \multicolumn{1}{c|}{0.48} & \multicolumn{1}{c|}{0.6} & \multicolumn{1}{c|}{49} \\ \hline
\multicolumn{1}{|c|}{LinearSVC} & \multicolumn{1}{c|}{28} & \multicolumn{1}{c|}{73.36} & \multicolumn{1}{c|}{4.12} & \multicolumn{1}{c|}{0.81} & \multicolumn{1}{c|}{0.53} & \multicolumn{1}{c|}{0.55} & \multicolumn{1}{c|}{0.52} & \multicolumn{1}{c|}{0.58} & \multicolumn{1}{c|}{102} & \multicolumn{1}{c|}{-6.63} & \multicolumn{1}{c|}{-0.21} & \multicolumn{1}{c|}{0.49} & \multicolumn{1}{c|}{0.5} & \multicolumn{1}{c|}{0.54} & \multicolumn{1}{c|}{0.48} & \multicolumn{1}{c|}{0.6} & \multicolumn{1}{c|}{49} \\ \hline
\multicolumn{1}{|c|}{LogisticRegression} & \multicolumn{1}{c|}{28} & \multicolumn{1}{c|}{97.44} & \multicolumn{1}{c|}{5.43} & \multicolumn{1}{c|}{0.87} & \multicolumn{1}{c|}{0.53} & \multicolumn{1}{c|}{0.59} & \multicolumn{1}{c|}{0.52} & \multicolumn{1}{c|}{0.69} & \multicolumn{1}{c|}{94} & \multicolumn{1}{c|}{-20.12} & \multicolumn{1}{c|}{-1.96} & \multicolumn{1}{c|}{0.34} & \multicolumn{1}{c|}{0.49} & \multicolumn{1}{c|}{0.52} & \multicolumn{1}{c|}{0.47} & \multicolumn{1}{c|}{0.58} & \multicolumn{1}{c|}{51} \\ \hline
\multicolumn{1}{|c|}{LogisticRegressionCV} & \multicolumn{1}{c|}{28} & \multicolumn{1}{c|}{111.04} & \multicolumn{1}{c|}{6.57} & \multicolumn{1}{c|}{0.81} & \multicolumn{1}{c|}{0.56} & \multicolumn{1}{c|}{0.68} & \multicolumn{1}{c|}{0.53} & \multicolumn{1}{c|}{0.96} & \multicolumn{1}{c|}{36} & \multicolumn{1}{c|}{6.81} & \multicolumn{1}{c|}{1.47} & \multicolumn{1}{c|}{0.32} & \multicolumn{1}{c|}{0.52} & \multicolumn{1}{c|}{0.63} & \multicolumn{1}{c|}{0.5} & \multicolumn{1}{c|}{0.86} & \multicolumn{1}{c|}{31} \\ \hline
\multicolumn{1}{|c|}{MLPClassifier} & \multicolumn{1}{c|}{28} & \multicolumn{1}{c|}{\cellcolor[HTML]{FFCCC9}\textbf{112.04}} & \multicolumn{1}{c|}{\cellcolor[HTML]{FFCCC9}\textbf{4.94}} & \multicolumn{1}{c|}{\cellcolor[HTML]{FFCCC9}\textbf{0.77}} & \multicolumn{1}{c|}{\cellcolor[HTML]{FFCCC9}\textbf{0.55}} & \multicolumn{1}{c|}{\cellcolor[HTML]{FFCCC9}\textbf{0.62}} & \multicolumn{1}{c|}{\cellcolor[HTML]{FFCCC9}\textbf{0.53}} & \multicolumn{1}{c|}{\cellcolor[HTML]{FFCCC9}\textbf{0.75}} & \multicolumn{1}{c|}{\cellcolor[HTML]{FFCCC9}\textbf{74}} & \multicolumn{1}{c|}{-28.13} & \multicolumn{1}{c|}{-3.23} & \multicolumn{1}{c|}{0.46} & \multicolumn{1}{c|}{0.43} & \multicolumn{1}{c|}{0.51} & \multicolumn{1}{c|}{0.44} & \multicolumn{1}{c|}{0.63} & \multicolumn{1}{c|}{41} \\ \hline
\multicolumn{1}{|c|}{PassiveAggressiveClassifier} & \multicolumn{1}{c|}{21} & \multicolumn{1}{c|}{83.33} & \multicolumn{1}{c|}{4.94} & \multicolumn{1}{c|}{0.84} & \multicolumn{1}{c|}{0.53} & \multicolumn{1}{c|}{0.57} & \multicolumn{1}{c|}{0.5} & \multicolumn{1}{c|}{0.66} & \multicolumn{1}{c|}{87} & \multicolumn{1}{c|}{-40.23} & \multicolumn{1}{c|}{-5.67} & \multicolumn{1}{c|}{0.84} & \multicolumn{1}{c|}{0.36} & \multicolumn{1}{c|}{0.41} & \multicolumn{1}{c|}{0.36} & \multicolumn{1}{c|}{0.47} & \multicolumn{1}{c|}{42} \\ \hline
\multicolumn{1}{|c|}{Perceptron} & \multicolumn{1}{c|}{28} & \multicolumn{1}{c|}{75.61} & \multicolumn{1}{c|}{4.84} & \multicolumn{1}{c|}{0.9} & \multicolumn{1}{c|}{0.58} & \multicolumn{1}{c|}{0.6} & \multicolumn{1}{c|}{0.57} & \multicolumn{1}{c|}{0.64} & \multicolumn{1}{c|}{90} & \multicolumn{1}{c|}{-55.87} & \multicolumn{1}{c|}{-8.62} & \multicolumn{1}{c|}{0.85} & \multicolumn{1}{c|}{0.37} & \multicolumn{1}{c|}{0.33} & \multicolumn{1}{c|}{0.33} & \multicolumn{1}{c|}{0.33} & \multicolumn{1}{c|}{35} \\ \hline
\multicolumn{1}{|c|}{QuadraticDiscriminantAnalysis} & \multicolumn{1}{c|}{21} & \multicolumn{1}{c|}{90.09} & \multicolumn{1}{c|}{12.36} & \multicolumn{1}{c|}{0.89} & \multicolumn{1}{c|}{0.53} & \multicolumn{1}{c|}{0.66} & \multicolumn{1}{c|}{0.5} & \multicolumn{1}{c|}{0.97} & \multicolumn{1}{c|}{28} & \multicolumn{1}{c|}{-3.98} & \multicolumn{1}{c|}{-0.16} & \multicolumn{1}{c|}{0.53} & \multicolumn{1}{c|}{0.51} & \multicolumn{1}{c|}{0.61} & \multicolumn{1}{c|}{0.49} & \multicolumn{1}{c|}{0.79} & \multicolumn{1}{c|}{29} \\ \hline
\multicolumn{1}{|c|}{RandomForestClassifier} & \multicolumn{1}{c|}{21} & \multicolumn{1}{c|}{87.75} & \multicolumn{1}{c|}{6.3} & \multicolumn{1}{c|}{0.79} & \multicolumn{1}{c|}{0.53} & \multicolumn{1}{c|}{0.65} & \multicolumn{1}{c|}{0.5} & \multicolumn{1}{c|}{0.92} & \multicolumn{1}{c|}{40} & \multicolumn{1}{c|}{\cellcolor[HTML]{9AFF99}\textbf{15.38}} & \multicolumn{1}{c|}{\cellcolor[HTML]{9AFF99}\textbf{8.68}} & \multicolumn{1}{c|}{\cellcolor[HTML]{9AFF99}\textbf{0.84}} & \multicolumn{1}{c|}{\cellcolor[HTML]{9AFF99}\textbf{0.52}} & \multicolumn{1}{c|}{\cellcolor[HTML]{9AFF99}\textbf{0.66}} & \multicolumn{1}{c|}{\cellcolor[HTML]{9AFF99}\textbf{0.5}} & \multicolumn{1}{c|}{\cellcolor[HTML]{9AFF99}\textbf{0.98}} & \multicolumn{1}{c|}{\cellcolor[HTML]{9AFF99}\textbf{10}} \\ \hline
\multicolumn{1}{|c|}{RidgeClassifier} & \multicolumn{1}{c|}{28} & \multicolumn{1}{c|}{94.36} & \multicolumn{1}{c|}{4.17} & \multicolumn{1}{c|}{0.78} & \multicolumn{1}{c|}{0.51} & \multicolumn{1}{c|}{0.59} & \multicolumn{1}{c|}{0.5} & \multicolumn{1}{c|}{0.71} & \multicolumn{1}{c|}{74} & \multicolumn{1}{c|}{-7.02} & \multicolumn{1}{c|}{-0.23} & \multicolumn{1}{c|}{0.35} & \multicolumn{1}{c|}{0.48} & \multicolumn{1}{c|}{0.53} & \multicolumn{1}{c|}{0.47} & \multicolumn{1}{c|}{0.63} & \multicolumn{1}{c|}{51} \\ \hline
\multicolumn{1}{|c|}{SGDClassifier} & \multicolumn{1}{c|}{28} & \multicolumn{1}{c|}{104.29} & \multicolumn{1}{c|}{5.16} & \multicolumn{1}{c|}{0.87} & \multicolumn{1}{c|}{0.52} & \multicolumn{1}{c|}{0.57} & \multicolumn{1}{c|}{0.51} & \multicolumn{1}{c|}{0.64} & \multicolumn{1}{c|}{90} & \multicolumn{1}{c|}{-15.64} & \multicolumn{1}{c|}{-1.4} & \multicolumn{1}{c|}{0.68} & \multicolumn{1}{c|}{0.49} & \multicolumn{1}{c|}{0.53} & \multicolumn{1}{c|}{0.47} & \multicolumn{1}{c|}{0.6} & \multicolumn{1}{c|}{49} \\ \hline
\multicolumn{1}{|c|}{SVC} & \multicolumn{1}{c|}{28} & \multicolumn{1}{c|}{106.92} & \multicolumn{1}{c|}{5.24} & \multicolumn{1}{c|}{0.81} & \multicolumn{1}{c|}{0.52} & \multicolumn{1}{c|}{0.63} & \multicolumn{1}{c|}{0.51} & \multicolumn{1}{c|}{0.84} & \multicolumn{1}{c|}{60} & \multicolumn{1}{c|}{-12.8} & \multicolumn{1}{c|}{-1.34} & \multicolumn{1}{c|}{0.85} & \multicolumn{1}{c|}{0.46} & \multicolumn{1}{c|}{0.56} & \multicolumn{1}{c|}{0.46} & \multicolumn{1}{c|}{0.72} & \multicolumn{1}{c|}{35} \\ \hline
\multicolumn{1}{|c|}{RadiusNeighborsClassifier} & \multicolumn{1}{c|}{1} & \multicolumn{1}{c|}{26.97} & \multicolumn{1}{c|}{13.22} & \multicolumn{1}{c|}{0.8} & \multicolumn{1}{c|}{0.46} & \multicolumn{1}{c|}{0.63} & \multicolumn{1}{c|}{0.46} & \multicolumn{1}{c|}{0.99} & \multicolumn{1}{c|}{6} & \multicolumn{1}{c|}{\cellcolor[HTML]{9AFF99}\textbf{12.78}} & \multicolumn{1}{c|}{\cellcolor[HTML]{9AFF99}\textbf{5.08}} & \multicolumn{1}{c|}{\cellcolor[HTML]{9AFF99}\textbf{0.08}} & \multicolumn{1}{c|}{\cellcolor[HTML]{9AFF99}\textbf{0.49}} & \multicolumn{1}{c|}{\cellcolor[HTML]{9AFF99}\textbf{0.65}} & \multicolumn{1}{c|}{\cellcolor[HTML]{9AFF99}\textbf{0.48}} & \multicolumn{1}{c|}{\cellcolor[HTML]{9AFF99}\textbf{0.98}} & \multicolumn{1}{c|}{\cellcolor[HTML]{9AFF99}\textbf{6}} \\ \hline
\multicolumn{1}{l}{} & \multicolumn{1}{l}{} & \multicolumn{1}{l}{} & \multicolumn{1}{l}{} & \multicolumn{1}{l}{} & \multicolumn{1}{l}{} & \multicolumn{1}{l}{} & \multicolumn{1}{l}{} & \multicolumn{1}{l}{} & \multicolumn{1}{l}{} & \multicolumn{1}{l}{} & \multicolumn{1}{l}{} & \multicolumn{1}{l}{} & \multicolumn{1}{l}{} & \multicolumn{1}{l}{} & \multicolumn{1}{l}{} & \multicolumn{1}{l}{} & \multicolumn{1}{l}{}
\end{tabular}%
}
\caption{Performance Metrics of Classifiers: A Comparative Analysis of Backtest and Forwardtest Results}
\label{tab:zt3_classifiers}
\end{table*}

\section{Results and Discussion}
\subsection{Evaluation Metrics}
In the domain of algorithmic trading, the performance of classifiers and regressors is quantified through a series of established metrics. Each metric provides unique insights into the model's predictive accuracy, risk management, and overall economic viability. Below is a detailed explanation of each metric employed in this study:

\begin{itemize}
    \item \textbf{Profit and Loss (PNL) Percentage:} This metric measures the total percentage gain or loss of a trading strategy over a specified period. It is calculated by summing up individual trade outcomes (profit or loss) and dividing by the total investment. A positive PNL indicates profitability, while a negative PNL suggests a loss.

    \item \textbf{Sharpe Ratio:} Named after Nobel laureate William F. Sharpe, this ratio is used to understand the return of an investment compared to its risk. It is calculated by subtracting the risk-free rate of return from the average return of the investment and dividing the result by the investment's standard deviation. A higher Sharpe Ratio indicates a more desirable risk-adjusted return \cite{sharpe1994sharpe}.

    \item \textbf{R-squared (R2):} R2 is a statistical measure that represents the proportion of the variance for a dependent variable that's explained by an independent variable or variables in a regression model. An R2 of 1 indicates that the regression predictions perfectly fit the data.

    \item \textbf{Accuracy:} In classification tasks, accuracy is the fraction of predictions our model got right, defined as the number of correct predictions divided by the total number of predictions. It is a useful metric when the classes in the dataset are nearly balanced.

    \item \textbf{F1 Score:} The F1 score is the harmonic mean of precision and recall and is particularly useful when the class distribution is imbalanced. It is calculated as 2 times the product of precision and recall divided by the sum of precision and recall.

    \item \textbf{Precision:} Precision is defined as the number of true positives divided by the number of true positives plus the number of false positives. It is a measure of a classifier's exactness. A high precision relates to a low false positive rate.

    \item \textbf{Recall:} Recall, also known as sensitivity or true positive rate, is the number of true positives divided by the number of true positives plus the number of false negatives. It is a measure of a classifier's completeness.

    \item \textbf{Mean Absolute Error (MAE):} For regression models, MAE is a metric that sums the absolute differences between predicted and actual values and then takes the average. It gives an idea of how wrong the predictions were in terms of an average amount.

    \item \textbf{Mean Squared Error (MSE):} MSE is the average of the squares of the errors of the predictions. It penalizes larger errors more than smaller ones, due to squaring each difference.

    \item \textbf{Root Mean Squared Error (RMSE):} RMSE is the square root of the mean of the squared errors. It is commonly used in regression analysis to verify experimental results, and like MSE, gives more weight to larger errors.

    \item \textbf{Number of Trades:} This metric indicates the count of trades executed based on the model’s recommendations. It can provide an understanding of the model's trading frequency and has implications for transaction costs and market liquidity.
\end{itemize}

These metrics collectively provide a holistic view of the models' performance, enabling us to not only assess the profitability and accuracy of predictions but also to gauge the risk and reliability of the trading strategies derived from the models.

These metrics were chosen to provide a comprehensive evaluation of the models' performance. PNL, Sharpe Ratio, and Number of Trades directly relate to the financial effectiveness of the models. In contrast, R2, Accuracy, F1 Score, Precision, Recall, MAE, MSE, and RMSE offer insights into the models' predictive accuracy and error characteristics. A combination of these metrics allows for a balanced assessment, considering both financial viability and statistical accuracy.

\begin{table*}[]

\centering
\resizebox{\textwidth}{!}{%
\begin{tabular}{ccccccccccccccccll}
\cline{1-16}
\multicolumn{1}{|c|}{} & \multicolumn{1}{c|}{} & \multicolumn{7}{c|}{\textbf{Backtest}} & \multicolumn{7}{c|}{\textbf{Forwardtest}} &  &  \\ \cline{3-16}
\multicolumn{1}{|c|}{\multirow{-2}{*}{\textbf{Regressor}}} & \multicolumn{1}{c|}{\multirow{-2}{*}{\textbf{\begin{tabular}[c]{@{}c@{}}Rolling \\ window\end{tabular}}}} & \multicolumn{1}{c|}{\textbf{\begin{tabular}[c]{@{}c@{}}PNL \\ (\%)\end{tabular}}} & \multicolumn{1}{c|}{\textbf{Sharpe}} & \multicolumn{1}{c|}{\textbf{R2}} & \multicolumn{1}{c|}{\textbf{MAE}} & \multicolumn{1}{c|}{\textbf{MSE}} & \multicolumn{1}{c|}{\textbf{RMSE}} & \multicolumn{1}{c|}{\textbf{\begin{tabular}[c]{@{}c@{}}No. of \\ Trades\end{tabular}}} & \multicolumn{1}{c|}{\textbf{\begin{tabular}[c]{@{}c@{}}PNL \\ (\%)\end{tabular}}} & \multicolumn{1}{c|}{\textbf{Sharpe}} & \multicolumn{1}{c|}{\textbf{R2}} & \multicolumn{1}{c|}{\textbf{MAE}} & \multicolumn{1}{c|}{\textbf{MSE}} & \multicolumn{1}{c|}{\textbf{RMSE}} & \multicolumn{1}{c|}{\textbf{\begin{tabular}[c]{@{}c@{}}No. of \\ Trades\end{tabular}}} &  &  \\ \cline{1-16}
\multicolumn{1}{|c|}{AdaBoostRegressor} & \multicolumn{1}{c|}{28} & \multicolumn{1}{c|}{94.69} & \multicolumn{1}{c|}{7.62} & \multicolumn{1}{c|}{0.79} & \multicolumn{1}{c|}{0.0183} & \multicolumn{1}{c|}{0.0007} & \multicolumn{1}{c|}{0.0255} & \multicolumn{1}{c|}{25} & \multicolumn{1}{c|}{9.68} & \multicolumn{1}{c|}{2.73} & \multicolumn{1}{c|}{0.6} & \multicolumn{1}{c|}{0.0117} & \multicolumn{1}{c|}{0.0004} & \multicolumn{1}{c|}{0.0198} & \multicolumn{1}{c|}{16} &  &  \\ \cline{1-16}
\multicolumn{1}{|c|}{BaggingRegressor} & \multicolumn{1}{c|}{21} & \multicolumn{1}{c|}{\cellcolor[HTML]{FFCCC9}\textbf{102.04}} & \multicolumn{1}{c|}{\cellcolor[HTML]{FFCCC9}\textbf{6.56}} & \multicolumn{1}{c|}{\cellcolor[HTML]{FFCCC9}\textbf{0.92}} & \multicolumn{1}{c|}{\cellcolor[HTML]{FFCCC9}\textbf{0.0179}} & \multicolumn{1}{c|}{\cellcolor[HTML]{FFCCC9}\textbf{0.0006}} & \multicolumn{1}{c|}{\cellcolor[HTML]{FFCCC9}\textbf{0.0251}} & \multicolumn{1}{c|}{\cellcolor[HTML]{FFCCC9}\textbf{88}} & \multicolumn{1}{c|}{11.01} & \multicolumn{1}{c|}{1.99} & \multicolumn{1}{c|}{0.01} & \multicolumn{1}{c|}{0.0128} & \multicolumn{1}{c|}{0.0004} & \multicolumn{1}{c|}{0.0205} & \multicolumn{1}{c|}{52} &  &  \\ \cline{1-16}
\multicolumn{1}{|c|}{DecisionTreeRegressor} & \multicolumn{1}{c|}{21} & \multicolumn{1}{c|}{97.31} & \multicolumn{1}{c|}{6.34} & \multicolumn{1}{c|}{0.92} & \multicolumn{1}{c|}{0.0181} & \multicolumn{1}{c|}{0.0006} & \multicolumn{1}{c|}{0.0252} & \multicolumn{1}{c|}{64} & \multicolumn{1}{c|}{13.41} & \multicolumn{1}{c|}{2.39} & \multicolumn{1}{c|}{0.1} & \multicolumn{1}{c|}{0.0122} & \multicolumn{1}{c|}{0.0004} & \multicolumn{1}{c|}{0.0198} & \multicolumn{1}{c|}{38} &  &  \\ \cline{1-16}
\multicolumn{1}{|c|}{ExtraTreeRegressor} & \multicolumn{1}{c|}{28} & \multicolumn{1}{c|}{\cellcolor[HTML]{FFCCC9}\textbf{101.03}} & \multicolumn{1}{c|}{\cellcolor[HTML]{FFCCC9}\textbf{4.92}} & \multicolumn{1}{c|}{\cellcolor[HTML]{FFCCC9}\textbf{0.7}} & \multicolumn{1}{c|}{\cellcolor[HTML]{FFCCC9}\textbf{0.0183}} & \multicolumn{1}{c|}{\cellcolor[HTML]{FFCCC9}\textbf{0.0007}} & \multicolumn{1}{c|}{\cellcolor[HTML]{FFCCC9}\textbf{0.0255}} & \multicolumn{1}{c|}{\cellcolor[HTML]{FFCCC9}\textbf{80}} & \multicolumn{1}{c|}{-6.84} & \multicolumn{1}{c|}{-0.38} & \multicolumn{1}{c|}{0.81} & \multicolumn{1}{c|}{0.0121} & \multicolumn{1}{c|}{0.0004} & \multicolumn{1}{c|}{0.0201} & \multicolumn{1}{c|}{39} &  &  \\ \cline{1-16}
\multicolumn{1}{|c|}{GaussianProcessRegressor} & \multicolumn{1}{c|}{28} & \multicolumn{1}{c|}{90.8} & \multicolumn{1}{c|}{4.89} & \multicolumn{1}{c|}{0.73} & \multicolumn{1}{c|}{0.0183} & \multicolumn{1}{c|}{0.0007} & \multicolumn{1}{c|}{0.0256} & \multicolumn{1}{c|}{20} & \multicolumn{1}{c|}{-0.05} & \multicolumn{1}{c|}{-1000} & \multicolumn{1}{c|}{0} & \multicolumn{1}{c|}{0.0118} & \multicolumn{1}{c|}{0.0004} & \multicolumn{1}{c|}{0.0198} & \multicolumn{1}{c|}{0} &  &  \\ \cline{1-16}
\multicolumn{1}{|c|}{KNeighborsRegressor} & \multicolumn{1}{c|}{28} & \multicolumn{1}{c|}{\cellcolor[HTML]{FFCCC9}\textbf{106.01}} & \multicolumn{1}{c|}{\cellcolor[HTML]{FFCCC9}\textbf{6.71}} & \multicolumn{1}{c|}{\cellcolor[HTML]{FFCCC9}\textbf{0.94}} & \multicolumn{1}{c|}{\cellcolor[HTML]{FFCCC9}\textbf{0.0186}} & \multicolumn{1}{c|}{\cellcolor[HTML]{FFCCC9}\textbf{0.0006}} & \multicolumn{1}{c|}{\cellcolor[HTML]{FFCCC9}\textbf{0.0255}} & \multicolumn{1}{c|}{\cellcolor[HTML]{FFCCC9}\textbf{76}} & \multicolumn{1}{c|}{11.62} & \multicolumn{1}{c|}{2.09} & \multicolumn{1}{c|}{0.13} & \multicolumn{1}{c|}{0.0133} & \multicolumn{1}{c|}{0.0004} & \multicolumn{1}{c|}{0.0204} & \multicolumn{1}{c|}{41} &  &  \\ \cline{1-16}
\multicolumn{1}{|c|}{LinearSVR} & \multicolumn{1}{c|}{21} & \multicolumn{1}{c|}{71.57} & \multicolumn{1}{c|}{4.7} & \multicolumn{1}{c|}{0.86} & \multicolumn{1}{c|}{0.0181} & \multicolumn{1}{c|}{0.0007} & \multicolumn{1}{c|}{0.0256} & \multicolumn{1}{c|}{93} & \multicolumn{1}{c|}{24.49} & \multicolumn{1}{c|}{4.13} & \multicolumn{1}{c|}{0.3} & \multicolumn{1}{c|}{0.0124} & \multicolumn{1}{c|}{0.0004} & \multicolumn{1}{c|}{0.0199} & \multicolumn{1}{c|}{38} &  &  \\ \cline{1-16}
\multicolumn{1}{|c|}{MLPRegressor} & \multicolumn{1}{c|}{28} & \multicolumn{1}{c|}{76.92} & \multicolumn{1}{c|}{4.6} & \multicolumn{1}{c|}{0.86} & \multicolumn{1}{c|}{0.1229} & \multicolumn{1}{c|}{0.0236} & \multicolumn{1}{c|}{0.1536} & \multicolumn{1}{c|}{88} & \multicolumn{1}{c|}{-20.99} & \multicolumn{1}{c|}{-2.96} & \multicolumn{1}{c|}{0.71} & \multicolumn{1}{c|}{0.229} & \multicolumn{1}{c|}{0.0768} & \multicolumn{1}{c|}{0.2771} & \multicolumn{1}{c|}{34} &  &  \\ \cline{1-16}
\multicolumn{1}{|c|}{RandomForestRegressor} & \multicolumn{1}{c|}{28} & \multicolumn{1}{c|}{84.01} & \multicolumn{1}{c|}{14.67} & \multicolumn{1}{c|}{0.91} & \multicolumn{1}{c|}{0.0183} & \multicolumn{1}{c|}{0.0007} & \multicolumn{1}{c|}{0.0257} & \multicolumn{1}{c|}{8} & \multicolumn{1}{c|}{3.38} & \multicolumn{1}{c|}{2.72} & \multicolumn{1}{c|}{0.01} & \multicolumn{1}{c|}{0.0117} & \multicolumn{1}{c|}{0.0004} & \multicolumn{1}{c|}{0.0198} & \multicolumn{1}{c|}{10} &  &  \\ \cline{1-16}
\multicolumn{1}{|c|}{Ridge} & \multicolumn{1}{c|}{21} & \multicolumn{1}{c|}{37.35} & \multicolumn{1}{c|}{2.42} & \multicolumn{1}{c|}{0.45} & \multicolumn{1}{c|}{0.0197} & \multicolumn{1}{c|}{0.0007} & \multicolumn{1}{c|}{0.0264} & \multicolumn{1}{c|}{84} & \multicolumn{1}{c|}{20.92} & \multicolumn{1}{c|}{3.76} & \multicolumn{1}{c|}{0.56} & \multicolumn{1}{c|}{0.0163} & \multicolumn{1}{c|}{0.0005} & \multicolumn{1}{c|}{0.0221} & \multicolumn{1}{c|}{45} &  &  \\ \cline{1-16}
\multicolumn{1}{|c|}{SGDRegressor} & \multicolumn{1}{c|}{28} & \multicolumn{1}{c|}{81.28} & \multicolumn{1}{c|}{5.06} & \multicolumn{1}{c|}{0.87} & \multicolumn{1}{c|}{0.0184} & \multicolumn{1}{c|}{0.0007} & \multicolumn{1}{c|}{0.0256} & \multicolumn{1}{c|}{63} & \multicolumn{1}{c|}{\cellcolor[HTML]{9AFF99}\textbf{34.01}} & \multicolumn{1}{c|}{\cellcolor[HTML]{9AFF99}\textbf{5.34}} & \multicolumn{1}{c|}{\cellcolor[HTML]{9AFF99}\textbf{0.8}} & \multicolumn{1}{c|}{\cellcolor[HTML]{9AFF99}\textbf{0.0117}} & \multicolumn{1}{c|}{\cellcolor[HTML]{9AFF99}\textbf{0.0004}} & \multicolumn{1}{c|}{\cellcolor[HTML]{9AFF99}\textbf{0.0195}} & \multicolumn{1}{c|}{\cellcolor[HTML]{9AFF99}\textbf{38}} &  &  \\ \cline{1-16}
\multicolumn{1}{|c|}{SVR} & \multicolumn{1}{c|}{7} & \multicolumn{1}{c|}{76.74} & \multicolumn{1}{c|}{4.86} & \multicolumn{1}{c|}{0.73} & \multicolumn{1}{c|}{0.0272} & \multicolumn{1}{c|}{0.0013} & \multicolumn{1}{c|}{0.0355} & \multicolumn{1}{c|}{81} & \multicolumn{1}{c|}{-24.45} & \multicolumn{1}{c|}{-2.81} & \multicolumn{1}{c|}{0.61} & \multicolumn{1}{c|}{0.0261} & \multicolumn{1}{c|}{0.0012} & \multicolumn{1}{c|}{0.0341} & \multicolumn{1}{c|}{50} &  &  \\ \cline{1-16}
\multicolumn{1}{|c|}{ARDRegression} & \multicolumn{1}{c|}{28} & \multicolumn{1}{c|}{76.33} & \multicolumn{1}{c|}{4.99} & \multicolumn{1}{c|}{0.8} & \multicolumn{1}{c|}{0.0183} & \multicolumn{1}{c|}{0.0007} & \multicolumn{1}{c|}{0.0257} & \multicolumn{1}{c|}{42} & \multicolumn{1}{c|}{17.54} & \multicolumn{1}{c|}{3.6} & \multicolumn{1}{c|}{0.12} & \multicolumn{1}{c|}{0.0119} & \multicolumn{1}{c|}{0.0004} & \multicolumn{1}{c|}{0.0199} & \multicolumn{1}{c|}{13} &  &  \\ \cline{1-16}
\multicolumn{1}{|c|}{BayesianRidge} & \multicolumn{1}{c|}{28} & \multicolumn{1}{c|}{59.04} & \multicolumn{1}{c|}{4.67} & \multicolumn{1}{c|}{0.85} & \multicolumn{1}{c|}{0.0185} & \multicolumn{1}{c|}{0.0007} & \multicolumn{1}{c|}{0.0256} & \multicolumn{1}{c|}{47} & \multicolumn{1}{c|}{15.2} & \multicolumn{1}{c|}{2.67} & \multicolumn{1}{c|}{0.09} & \multicolumn{1}{c|}{0.0118} & \multicolumn{1}{c|}{0.0004} & \multicolumn{1}{c|}{0.0195} & \multicolumn{1}{c|}{29} &  &  \\ \cline{1-16}
\multicolumn{1}{|c|}{GradientBoostingRegressor} & \multicolumn{1}{c|}{28} & \multicolumn{1}{c|}{80.81} & \multicolumn{1}{c|}{6.44} & \multicolumn{1}{c|}{0.83} & \multicolumn{1}{c|}{0.0185} & \multicolumn{1}{c|}{0.0006} & \multicolumn{1}{c|}{0.0254} & \multicolumn{1}{c|}{30} & \multicolumn{1}{c|}{0.43} & \multicolumn{1}{c|}{2.36} & \multicolumn{1}{c|}{0.05} & \multicolumn{1}{c|}{0.0123} & \multicolumn{1}{c|}{0.0004} & \multicolumn{1}{c|}{0.02} & \multicolumn{1}{c|}{6} &  &  \\ \cline{1-16}
\multicolumn{1}{|c|}{Lars} & \multicolumn{1}{c|}{21} & \multicolumn{1}{c|}{48.78} & \multicolumn{1}{c|}{3.18} & \multicolumn{1}{c|}{0.74} & \multicolumn{1}{c|}{0.0461} & \multicolumn{1}{c|}{0.004} & \multicolumn{1}{c|}{0.063} & \multicolumn{1}{c|}{105} & \multicolumn{1}{c|}{\cellcolor[HTML]{9AFF99}\textbf{31.69}} & \multicolumn{1}{c|}{\cellcolor[HTML]{9AFF99}\textbf{4.88}} & \multicolumn{1}{c|}{\cellcolor[HTML]{9AFF99}\textbf{0.66}} & \multicolumn{1}{c|}{\cellcolor[HTML]{9AFF99}\textbf{0.0525}} & \multicolumn{1}{c|}{\cellcolor[HTML]{9AFF99}\textbf{0.0051}} & \multicolumn{1}{c|}{\cellcolor[HTML]{9AFF99}\textbf{0.0711}} & \multicolumn{1}{c|}{\cellcolor[HTML]{9AFF99}\textbf{42}} &  &  \\ \cline{1-16}
\multicolumn{1}{|c|}{LinearRegression} & \multicolumn{1}{c|}{28} & \multicolumn{1}{c|}{48.98} & \multicolumn{1}{c|}{3.14} & \multicolumn{1}{c|}{0.5} & \multicolumn{1}{c|}{0.1156} & \multicolumn{1}{c|}{0.0211} & \multicolumn{1}{c|}{0.1452} & \multicolumn{1}{c|}{79} & \multicolumn{1}{c|}{\cellcolor[HTML]{9AFF99}\textbf{27.64}} & \multicolumn{1}{c|}{\cellcolor[HTML]{9AFF99}\textbf{4.57}} & \multicolumn{1}{c|}{\cellcolor[HTML]{9AFF99}\textbf{0.66}} & \multicolumn{1}{c|}{\cellcolor[HTML]{9AFF99}\textbf{0.1197}} & \multicolumn{1}{c|}{\cellcolor[HTML]{9AFF99}\textbf{0.0231}} & \multicolumn{1}{c|}{\cellcolor[HTML]{9AFF99}\textbf{0.1519}} & \multicolumn{1}{c|}{\cellcolor[HTML]{9AFF99}\textbf{38}} &  &  \\ \cline{1-16}
\multicolumn{1}{|c|}{RANSACRegressor} & \multicolumn{1}{c|}{21} & \multicolumn{1}{c|}{47.17} & \multicolumn{1}{c|}{2.97} & \multicolumn{1}{c|}{0.57} & \multicolumn{1}{c|}{0.1399} & \multicolumn{1}{c|}{0.0316} & \multicolumn{1}{c|}{0.1777} & \multicolumn{1}{c|}{97} & \multicolumn{1}{c|}{-6.12} & \multicolumn{1}{c|}{-0.16} & \multicolumn{1}{c|}{0.08} & \multicolumn{1}{c|}{0.1487} & \multicolumn{1}{c|}{0.0345} & \multicolumn{1}{c|}{0.1857} & \multicolumn{1}{c|}{49} &  &  \\ \cline{1-16}
\multicolumn{1}{|c|}{TheilSenRegressor} & \multicolumn{1}{c|}{7} & \multicolumn{1}{c|}{81.96} & \multicolumn{1}{c|}{4.45} & \multicolumn{1}{c|}{0.75} & \multicolumn{1}{c|}{0.1429} & \multicolumn{1}{c|}{0.0474} & \multicolumn{1}{c|}{0.2176} & \multicolumn{1}{c|}{70} & \multicolumn{1}{c|}{-7.02} & \multicolumn{1}{c|}{-0.42} & \multicolumn{1}{c|}{0.73} & \multicolumn{1}{c|}{0.1724} & \multicolumn{1}{c|}{0.0564} & \multicolumn{1}{c|}{0.2374} & \multicolumn{1}{c|}{31} &  &  \\ \cline{1-16}
\multicolumn{1}{|c|}{RadiusNeighborsRegressor} & \multicolumn{1}{c|}{1} & \multicolumn{1}{c|}{42.09} & \multicolumn{1}{c|}{3.65} & \multicolumn{1}{c|}{0.67} & \multicolumn{1}{c|}{0.0175} & \multicolumn{1}{c|}{0.0007} & \multicolumn{1}{c|}{0.0255} & \multicolumn{1}{c|}{42} & \multicolumn{1}{c|}{1.69} & \multicolumn{1}{c|}{1.12} & \multicolumn{1}{c|}{0.16} & \multicolumn{1}{c|}{0.0126} & \multicolumn{1}{c|}{0.0004} & \multicolumn{1}{c|}{0.0211} & \multicolumn{1}{c|}{20} &  &  \\ \cline{1-16}
\end{tabular}%
}
\caption{Performance Metrics of Regressors: Evaluating Predictive Strength Across Market Conditions}
\label{tab:zt3_regressors}
\end{table*}

\subsection{Classifier Results Interpretation}
\label{sec:classifier_results_interpretation}

Table~\ref{tab:zt3_classifiers} provides a quantitative evaluation of classifier models over two distinct phases: backtesting and forward testing. The performance of each classifier is contextualized by a set of metrics, and the rolling window sizes are instrumental in capturing temporal market dynamics. The top-performing models in each phase are highlighted, indicating their superior ability to navigate the complexities of market prediction.

\subsubsection{Backtest Insights}
The backtest phase reveals the intrinsic strength of the classifiers when applied to historical data. For instance, the highlighted BaggingClassifier, with a rolling window of 28 days, achieved an exceptional PNL, suggesting that its ensemble approach is particularly suited to grasp long-term trends. Conversely, the BernoulliNB classifier demonstrates a high degree of precision in the shorter rolling window of 21 days, indicating its potential effectiveness in short-term market movement prediction. The MLPClassifier's balanced metrics, particularly its F1 score, suggest a well-tuned model that avoids overfitting, evidenced by its ability to maintain high precision and recall.

\subsubsection{Forward Test Observations}
The forward testing phase is critical for assessing the real-world applicability of the classifiers. The Random Forest Classifier, which maintained a consistent performance across both phases, indicates not just a strong fit to the data but also adaptability to evolving market conditions. The sharp increase in Sharpe Ratio for the Quadratic Discriminant Analysis and RadiusNeighborsClassifier from backtest to forward test underscores their potential for yielding profitable strategies when applied in real-time, despite their less impressive backtest PNL. These results underscore the importance of evaluating models on unseen data to gauge their practical utility.

\subsubsection{Rolling Window and Model Responsiveness}
The varying rolling window sizes play a significant role in the classifiers' ability to capture different market conditions. Larger windows may allow classifiers to integrate longer-term trends into their predictions, which can be crucial for capturing macroeconomic movements that affect asset prices. Smaller windows, on the other hand, may enable classifiers to react more quickly to short-term market volatility, which could be advantageous in rapidly changing trading environments.

\subsubsection{Interpreting the Discrepancies Between Backtest and Forward Test Results}
The highlighted models exhibit varied performances when transitioning from backtest to forward test environments. Such discrepancies may stem from overfitting to historical data patterns that do not extrapolate well into future market states. The BaggingClassifier, while performing optimally in backtesting, shows a decrease in PNL during forward testing. This could indicate a model finely tuned to past conditions but less adaptable to unforeseen market shifts. In contrast, the Random Forest Classifier demonstrates robustness, with a more consistent PNL, suggesting a model that captures underlying market drivers that persist over time.

\subsubsection{Assessing Model Robustness and Economic Significance}
Robustness in financial models is demonstrated by consistent performance across both backtesting and forward testing. Economic significance, however, is derived from the model's ability to produce actionable insights leading to profitable trades. The BernoulliNB classifier, for instance, maintains a high PNL in both phases, reinforcing its potential for real-world application. The Sharpe Ratios, especially in forward test results, reflect the models' capabilities to deliver returns above the risk-free rate, which is crucial for long-term investment strategies.

\begin{figure*}[ht]
\centering
\includegraphics[width=\textwidth]{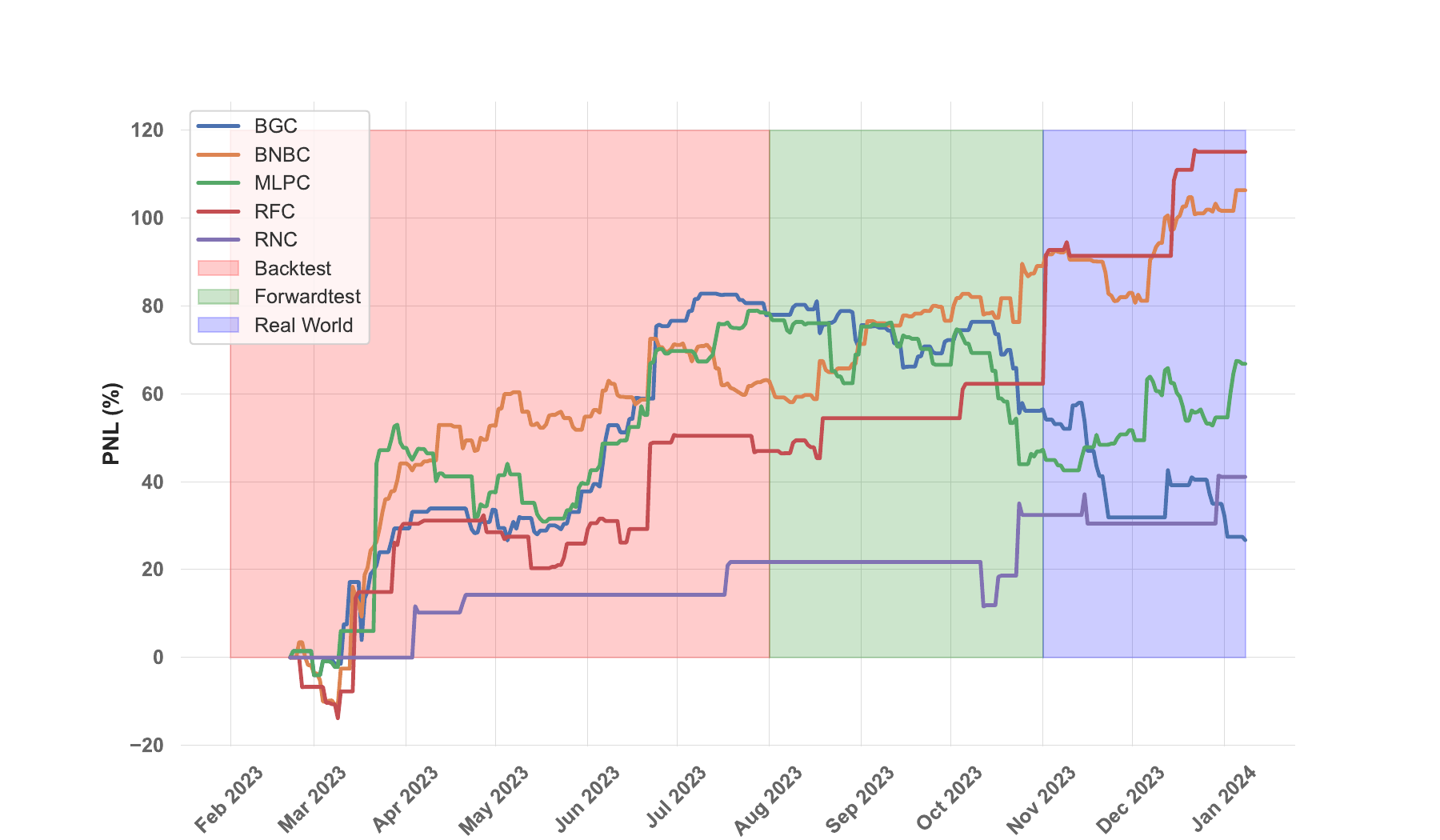}
\caption{Profit and Loss (PNL) Trajectories of Top Classifiers in Real-World Trading Scenarios}
\label{fig:real_world_pnl}
\end{figure*}

\subsection{Regressor Results Interpretation}
\label{sec:regressor_results_interpretation}
In parallel, Table~\ref{tab:zt3_regressors} lays out the regressors' performance, where the highlighted models exhibit noteworthy predictive power. Each regressor is scrutinized under metrics that collectively portray its predictive accuracy and economic impact.

\subsubsection{Backtest Insights}
During the backtest period, the SGDRegressor distinguished itself with a notable PNL and the highest Sharpe Ratio, suggesting effective risk management combined with profitability. This is further corroborated by its relatively high R2 value, reflecting the model's capability to capture the variance in price movement effectively. The GradientBoostingRegressor and Lars, both highlighted for their substantial PNL, also demonstrate solid R2 scores, which points to their models' good explanatory power.

\subsubsection{Forward Test Observations}
Transitioning to forward testing, the SGDRegressor maintains a strong performance, indicating robustness and potential for real-world application. The Lars regressor shows an increase in both PNL and Sharpe Ratio, suggesting that its simpler, linear approach is well-suited for the forward test market conditions. The consistency in the performance of the RadiusNeighborsRegressor, with minimal trades, accentuates its precision in trade selection, which is vital for strategies aiming to minimize transaction costs.

\subsubsection{Rolling Window and Model Predictive Dynamics}
The regressors' results highlight the significance of selecting an appropriate rolling window size, which directly influences their ability to assimilate and predict based on the market's historical data. The rolling window's impact is evident in the models' varied performance across the two testing phases, with different window lengths aligning with specific market behaviors that the models have learned to predict.

\subsubsection{Analysis of Regressor Robustness}
The robustness of regressors is evaluated through their ability to maintain predictive accuracy from backtesting to live-market forward testing. The SGDRegressor, with its high Sharpe Ratio and consistent PNL, exemplifies a model with a stable foundation, likely to withstand market volatilities. The Rolling Window's significance is evident in the models' ability to incorporate relevant market data into their predictive framework, with longer windows capturing more extensive market trends.

\subsubsection{Economic and Predictive Implications}
The economic implications of the regressors' performance are multifaceted. A high PNL is desirable but must be coupled with low predictive error metrics, such as MAE and RMSE, to be economically significant. The Lars model, for instance, illustrates this with an improved Sharpe Ratio and a lower RMSE in forward testing, suggesting a model that not only forecasts accurately but does so with economic prudence.

\subsection{Closing Evaluation}
The detailed analysis of classifiers and regressors underscores the multifaceted nature of financial prediction. The highlighted models in the tables provide a benchmark for what can be achieved with careful tuning and selection of rolling windows. These results emphasize the necessity of a comprehensive evaluation framework that incorporates a variety of performance metrics to assess model efficacy thoroughly. The findings from the backtest and forward test phases offer invaluable insights for developing resilient trading strategies capable of adapting to the ever-evolving patterns of financial markets.

\subsection{Hyperparameter Optimization: Tuning for Peak Performance}
In the quest for optimal model performance, hyperparameter optimization serves as the fine-tuning process that can make or break the predictive power of classifiers and regressors. The hyperparameter optimization for classifiers was meticulously performed using advanced techniques that explored the depth and breadth of the parameter space, striking a balance between model complexity and generalization capability. The regressors underwent a similar process, with each model's unique parameters adjusted to navigate the intricate landscape of financial time series forecasting. This iterative and methodical approach ensured that the final model configurations were not just suited to historical patterns but were also robust and flexible enough to adapt to new, unseen market data.

\begin{figure*}[ht]
\centering
\includegraphics[width=\textwidth]{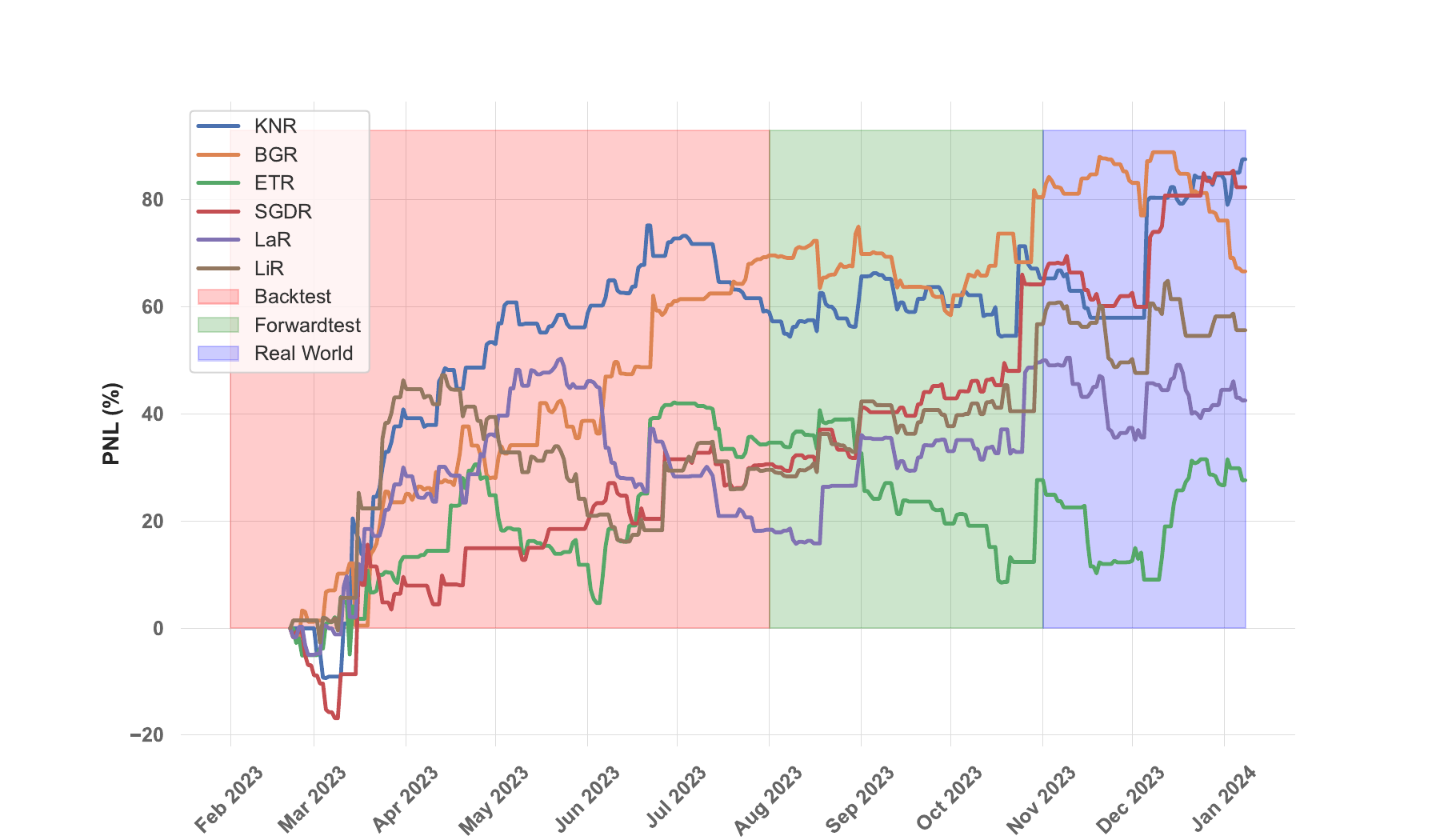}
\caption{Real-World Profit and Loss (PNL) Performance of Top Regressors}
\label{fig:regressors_real_world_pnl}
\end{figure*}

\subsection{Analysis of Top Models on Real-World Data}
\label{sec:analysis_top_models_real_world}
An empirical evaluation of the top-performing classifiers was conducted to assess their ability to generalize beyond backtesting and forward testing scenarios. This analysis is crucial to determine the models' viability in live-market conditions, where unpredictability and external factors play a significant role.

\subsubsection{Interpreting Classifier Performance in the Real World}
\label{sec:interpretation_classifier_performance_real_world}
Figure~\ref{fig:real_world_pnl} illustrates the Profit and Loss (PNL) trajectories of the top classifiers over a timeline that spans backtesting, forward testing, and into the real-world application phase. Each line represents the PNL progression of a model, providing insights into their performance stability and adaptability to real market conditions.

The shaded areas—red for backtesting, green for forward testing, and blue for the real-world phase—contextualize the timeline of each model's deployment. Across the transition from controlled testing environments to the real world, the following observations are made:

\begin{itemize}
    \item \textbf{Consistency of Performance:} The models that maintain a steady trajectory from backtesting through to real-world trading, such as the Random Forest Classifier (RFC), indicate a strong ability to adapt to evolving market conditions without overfitting to historical data.
    \item \textbf{Adaptability to Market Shifts:} Some models, like the Multi-Layer Perceptron Classifier (MLPC), show resilience in the face of market volatility, as evidenced by their PNL performance remaining robust or improving when transitioning to real-world trading.
    \item \textbf{Real-World Viability:} The Bagging Classifier (BGC) and BernoulliNB Classifier (BNBC) demonstrate significant real-world viability, highlighted by their sustained PNL levels in the live market phase. This suggests that these models have captured fundamental market drivers that are applicable in ongoing trading.
    \item \textbf{Economic Significance:} The Ridge Classifier (RNC), while showing a dip in the forward test phase, recovers in the real-world application, pointing to economic strategies embedded within the model that may only become evident under actual market pressures.
    \item \textbf{Volatility and Risk Management:} The volatility in the PNL trajectories for some classifiers indicates the varying risk profiles and the models' sensitivity to market fluctuations. Effective risk management strategies are imperative for these models to ensure that high volatility does not erode profitability.
\end{itemize}

The detailed visualization of PNL trajectories in Figure~\ref{fig:real_world_pnl} serves as a testament to the models' capabilities and provides a predictive lens through which investors can gauge the potential success of deploying these models in live trading scenarios. The analysis confirms that while backtest and forwardtest performances are indicative, the ultimate test for any trading model lies in its real-world application.

\subsubsection{Real-World Performance of Regressors}
\label{sec:real_world_performance_regressors}
Figure~\ref{fig:regressors_real_world_pnl} presents the PNL performance of selected regressor models as they transition from the controlled environments of backtesting and forward testing into actual market deployment. The PNL trajectories provide a longitudinal view of each model's ability to navigate and capitalize on real market trends.

The shaded regions represent different evaluation phases: backtesting (red), forward testing (green), and the real-world trading period (blue). The regressors' performance trends across these phases offer a multifaceted perspective on their predictive capabilities and economic utility:

\begin{itemize}
    \item \textbf{KNeighborsRegressor (KNR)} displays a relatively stable PNL during backtesting, which declines during forward testing but shows recovery in real-world conditions. This pattern suggests a sensitivity to market conditions that may require adaptive parameter adjustments or dynamic feature selection to maintain profitability.

    \item \textbf{BaggingRegressor (BGR)} and \textbf{ExtraTreesRegressor (ETR)} both demonstrate high PNL in the backtest phase, with the BGR maintaining this performance in the forward test phase, indicating a robust model less prone to overfitting and capable of capturing persistent market signals.

    \item \textbf{Stochastic Gradient Descent Regressor (SGDR)} shows a consistent increase in PNL across all phases, highlighting its strength in adapting to new data. Its performance in the real-world phase, in particular, underscores the potential of SGD-based models for financial time series forecasting.

    \item \textbf{Lasso Regression (LaR)} and \textbf{Linear Regression (LiR)} exhibit significant PNL volatility post-backtesting. The divergence in their PNL during the real-world phase could reflect their varying degrees of regularization and feature weighting, which impact their ability to handle non-stationary market data.

    \item \textbf{Performance Fluctuations:} The fluctuations and drops in PNL for some models from backtesting to real-world application highlight the challenges of model generalization and the impact of market volatility. These variations call for ongoing model recalibration and robust risk management strategies to mitigate potential drawdowns.
\end{itemize}

The PNL trajectories in Figure~\ref{fig:regressors_real_world_pnl} underscore the importance of rigorous model evaluation. Models that demonstrate resilience and adaptability in forward testing are more likely to perform well in real-world trading, but the ultimate litmus test for any trading strategy is its ability to sustain profitability in the live market. This graph illustrates not only the successes but also the limitations of the tested regressors, guiding future model refinement and the development of adaptive trading systems.

\section{Conclusion} 
This study evaluated the performance of 41 machine learning models, comprising 21 classifiers and 20 regressors, for Bitcoin price prediction in algorithmic trading. Through rigorous backtesting, forward testing, and real-world testing, we identified that models like Random Forest and Stochastic Gradient Descent exhibit superior performance in terms of profit and risk management. The integration of both machine learning metrics (e.g., Mean Absolute Error, Root Mean Squared Error) and trading metrics (e.g., Profit and Loss percentage, Sharpe Ratio) provided a comprehensive assessment of model performance.

Our findings underscore the necessity for a multi-faceted evaluation approach to ensure the practical utility of trading models. Many models that performed well in backtesting did not translate effectively to forward tests and real-world scenarios, highlighting the limitations of relying solely on backtesting. By incorporating economic indicators and considering practical trading constraints, our study offers a robust and practical solution for Bitcoin price prediction and trading.

Future research should extend these findings to other cryptocurrencies and investigate the impact of different economic indicators on model performance. Additionally, exploring emerging machine learning techniques can further enhance predictive accuracy and trading effectiveness. This study provides valuable insights for traders and researchers aiming to leverage machine learning for more strategic and profitable cryptocurrency trading. Future work will also focus on refining our multi-faceted evaluation framework and exploring its application in different market conditions to further validate and improve the robustness of trading models.

\bibliographystyle{IEEEtran} 
\bibliography{References}
\end{document}